# Kinetics of dilute polymers with the viscous memory

A.V. Zatovsky[a] and V. Lisy[b]

[a]Department of Theoretical Physics, I.I. Mechnikov Odessa National University,
2, Dvoryanskaya Str., 65026 Odessa, Ukraine[+]

[b]Department of Biophysics, P.J. Safarik University, Jesenna 5, 04154 Kosice, Slovakia[*]

The Zimm equation for the position vector of a polymer segment is generalized taking into account the effect of viscous memory. The Oseen tensor is built on the basis of the nonstationary Navier-Stokes equation. After the preliminary averaging of the tensor a non-Markovian equation for the time correlation function of the Fourier components of the segment position is derived. The viscous memory essentially affects the long-time asymptote of the correlation function of the Fourier amplitudes of the chain fragments – the time behavior is of a fractional power character instead of the traditional exponential. The relaxation time and the diffusion coefficient of the macromolecule as a whole are the same as in the Zimm model.

Building the theory of the kinetic phenomena in dilute solutions of polymers, the macromolecule is often represented by a set of beads jointed into the chain. The dynamics of such a polymer molecule is modeled by the Brownian motion of the beads. The analysis of the dynamic properties of polymer solutions in the Zimm model [1] is based on the account for the Stokes force that acts on the beads, and on the perturbation of the velocity field of the solvent due to the motion of neighboring beads, so that the hydrodynamic interaction is described by the Oseen tensor. The evaluation of this tensor is within the Zimm model carried out with the use of linearized Navier-Stokes equations for the stationary motion of the solvent. In the approximation of preliminary averaging of the Oseen tensor in the Zimm equation this leads to the exponential time relaxation [1,2] of the correlation functions of the normal modes of the polymer chain, that corresponds to the Markovian process of Brownian walking [3]. In the general case, on every particle of the model chain a force acts that depends on the relative velocity of the particle and the solvent in all the foregoing moments of time (the Boussinesq friction force), and the Oseen tensor should take into account the viscous aftereffect. For non-interacting Brownian particles this leads to an essentially different, algebraic, not the exponential asymptote of the time correlation function of dynamic variables [4-6], that reflects collective properties of the correlation functions of the molecules of the liquid [7].

We have generalized the Zimm equation for the vector of the position of arbitrary polymer segment taking into account the effects of viscous memory. First of all this concerns the evaluation of the Oseen tensor on the basis of the nonstationary Navier-Stokes equation. After the procedure of preliminary averaging of the tensor, a non-Markovian equation is obtained for the time correlation function of the Fourier components of the chain segment position. The viscous memory significantly affects the long-time asymptote of the Fourier components of the fragments of polymer chain – it has a fractional power character instead of the traditional

---

[+] E-mail: avz@dtp.odessa.ua
[*] E-mail: lisy@kosice.upjs.sk

exponential one. The maximum relaxation time and the diffusion coefficient of the macromolecule as a whole are the same as in the Zimm model.

The equation of the Brownian motion of the *n*th segment of the polymer chain has a form

$$M \frac{d^2 \vec{x}_n}{dt^2} = -\xi \left( \frac{d\vec{x}_n}{dt} - \vec{v}(\vec{x}_n) \right) - \frac{\partial u}{\partial \vec{x}_n} + \vec{f}_n . \tag{1}$$

Here $\vec{x}_n$ is the position vector of the chain segment (the bead) of the mass $M$. The first term on the right-hand side is the Stokes friction force on the bead. It takes into account that the bead is carried by the velocity field of the solvent in the point of the bead, due to the motion of neighboring segments of the polymer chain (the hydrodynamic interaction). The interaction between the segments is described by the effective potential [2]

$$u = \frac{3k_B T}{2a^2} \sum_{n=2}^{N} (\vec{x}_n - \vec{x}_{n-1})^2 \tag{2}$$

($a$ is the mean-square distance between the neighboring beads along the chain), and $\vec{f}_n$ is the Gaussian random force with the zero mean value. In the simplest case the Brownian motion is considered as a Markovian random process, $\xi = 6\pi b \eta$ ($\eta$ is the viscosity of the solvent and $b$ is the radius of the spherical bead), and the flow excited by the beads is stationary. The account for the viscous aftereffect in an incompressible solvent leads to the integro-differential Volterra equation and in the Fourier representation instead of $\xi$ the following expression has to be used [4]:

$$\xi^\omega = 6\pi b \eta \left[ 1 + \chi b + \frac{1}{9} (\chi b)^2 \right], \qquad \chi = \sqrt{-i\omega/\nu} . \tag{3}$$

Here $\text{Re}\chi > 0$ and $\nu = \eta/\rho$ is the kinematic viscosity of the solvent. Since the solvent is an incompressible viscous fluid and the flow excited by the polymer chain is slow, the Navier-Stokes equations can be written in the linearized form,

$$\rho \frac{\partial \vec{v}}{\partial t} = -\nabla p + \eta \Delta \vec{v} + \vec{\varphi}, \quad \text{div } \vec{v} = 0, \tag{4}$$

where $p$ is the pressure and $\vec{\varphi}$ is the external force per unit volume, acting on the solvent near $\vec{x}_n$. The external force is expressed as follows [2,8]:

$$\vec{\varphi}(\vec{x}, t) = \sum_n \left( -\frac{\partial u}{\partial \vec{x}_n} + \vec{f}_n \right) \delta(\vec{x} - \vec{x}_n). \tag{5}$$

After the Fourier transformation in the time the solution of Eqs. (4) are easily found in the form of projections on the Cartesian axes,

$$v_\alpha^\omega(\vec{r}) = \int d\vec{r}' \, H_{\alpha\beta}^\omega (\vec{r} - \vec{r}') \varphi_\beta^\omega(\vec{r}') . \tag{6}$$

Here the Fourier representation of the nonstationary Oseen mobility tensor is introduced:

$$H^{\omega}_{\alpha\beta}(\vec{r}) = A\delta_{\alpha\beta} + B\frac{r_\alpha r_\beta}{r^2},$$

$$A = \frac{1}{8\pi\eta r}\left[e^{-y} - y\left(\frac{1-e^{-y}}{y}\right)''\right], \quad B = \frac{1}{8\pi\eta r}\left[e^{-y} + 3y\left(\frac{1-e^{-y}}{y}\right)''\right], \quad y = r\chi, \quad (7)$$

The prime denotes the differentiation with respect to $y$. In the particular case $\omega = 0$

$$H^{0}_{\alpha\beta}(\vec{r}) = \frac{1}{8\pi\eta r}\left(\delta_{\alpha\beta} + \frac{r_\alpha r_\beta}{r^2}\right),$$

and it coincides with the result by Zimm. Now the solution of the problem of perturbation of the velocity field due to the motion of the polymer chain can be written in the form

$$v_\alpha(\vec{x},t) = \sum_n \int_{-\infty}^{t} dt' H_{\alpha\beta}(\vec{x}-\vec{x}_n, t-t')\left(-\frac{\partial u(t')}{\partial \vec{x}_{n\beta}} + f_{n\beta}(t')\right). \quad (8)$$

In the continuum limit with respect to the discrete variable, the new equation of motion for the $n$th segment has the same form as in the Zimm model but in the Fourier representation:

$$-i\omega x^{\omega}_\alpha(n) = \int_0^N dm\, \mathrm{H}^{\omega}_{\alpha\beta nm}\left[\frac{3k_BT}{a^2}\frac{\partial^2 x^{\omega}_\beta(m)}{\partial m^2} + f^{\omega}_\beta(n)\right], \quad (9)$$

$$\mathrm{H}^{\omega}_{\alpha\beta nm} = \frac{\xi^{\omega}}{-i\omega M + \xi^{\omega}} H^{\omega}_{\alpha\beta}(\vec{x}_n - \vec{x}_m), \quad m \neq n, \quad \mathrm{H}^{\omega}_{\alpha\beta nm} = \frac{1}{-i\omega M + \xi^{\omega}}\delta_{\alpha\beta}, \quad m = n. \quad (10)$$

Due to the dependence of the Oseen tensor on the difference $\vec{x}_n - \vec{x}_m$, Eq. (9) is nonlinear and is hardly solvable analytically. We shall use the approximation of preliminary averaging replacing the expressions (10) by their averages over the equilibrium distribution function. Restricting ourselves to the states close to equilibrium, the pair distribution function is [8]

$$P(\vec{x}_n - \vec{x}_m) = \left(\frac{3}{2\pi a^2 |m-n|}\right)^{3/2} \exp\left\{-\frac{3}{2a^2}\frac{(\vec{x}_n - \vec{x}_m)^2}{|m-n|}\right\},$$

so that for $\langle H^{\omega}_{\alpha\beta}(\vec{x}_m - \vec{x}_n)\rangle_0$ we have

$$\langle H^{\omega}_{\alpha\beta}\rangle_0 = \left(\frac{3}{2\pi a^2 |m-n|}\right)^{3/2} \int d\vec{r}\, \exp\left[-\frac{3}{2a^2}\frac{r^2}{|m-n|}\right]\left(A(r)\delta_{\alpha\beta} + B(r)\frac{r_\alpha r_\beta}{r^2}\right) = \frac{\delta_{\alpha\beta}}{6\pi\eta}\left\langle\frac{e^{-\chi r}}{r}\right\rangle_0. \quad (11)$$

Taking into account Eq. (11), we search for the solution of Eq. (9) in the form of a superposition of the modes of displacements (the Fourier series in the variable $n$),

$$\vec{x}_n^\omega = \vec{y}_0^\omega + 2\sum_{p=1}^{\infty} \vec{y}_p^\omega \cos\frac{\pi p n}{N}, \qquad \vec{y}_p^\omega = \frac{1}{N}\int_0^N dn \cos\frac{\pi p n}{N}\, \vec{x}_n^\omega, \qquad p = 0, 1, 2, \ldots \tag{12}$$

From Eqs. (9) and (11) we then have

$$-i\omega\, y_{p\alpha}^\omega = \sum_{q=0}^{\infty} h_{\alpha\beta pq}^\omega \left(-\frac{6\pi^2 k_B T q^2}{Na^2} y_{q\beta}^\omega + f_{q\beta}^\omega\right), \tag{13}$$

$$h_{\alpha\beta pq}^\omega = \frac{1}{N^2}\int_0^N dn \int_0^N dm \cos\frac{\pi p n}{N} \cos\frac{\pi q m}{N} \langle \mathrm{H}_{\alpha\beta nm}^\omega \rangle_0, \tag{14}$$

and $\vec{f}_q^\omega$ is determined by the same transformation as in the second of Equations (12). The Oseen tensor after the averaging depends only on the difference $m-n$ and is diagonal with respect to the Cartesian indices, that is $\sim \delta_{\alpha\beta}$. At large values of $q$ the matrix $h_{\alpha\beta pq}$ is practically diagonal with respect to the indices of Fourier transformation, i.e., it is $\sim \delta_{pq}$. The prove of this statement does not differ from the case $\omega = 0$ [2]. When $p$ and $q$ are of the order of unity, the nondiagonal elements are also small compared with the diagonal ones. That is why the major terms of the matrix (14) are diagonal, so that after the integration we obtain

$$h_{\alpha\beta pq}^\omega = \frac{\delta_{\alpha\beta}\delta_{pq}}{\lambda_q^\omega},\ (q \neq 0),\quad \frac{1}{\lambda_q^\omega} = \frac{1}{\pi\eta a\sqrt{3\pi N q}}\frac{\xi^\omega}{-i\omega M + \xi^\omega}\frac{1+\chi_q}{1+(1+\chi_q)^2},\quad \chi_q = \sqrt{\frac{-i\omega N}{3\pi\nu q}}\, a. \tag{15}$$

When we neglect here the mass of the moving bead, setting $M = 0$, and the viscous aftereffect, the Zimm result is obtained. At $q = 0$ we find directly from the definition (14)

$$\lambda_0^\omega = 3\eta a\sqrt{6\pi^3 N}\,\frac{\xi^\omega - i\omega M}{4\xi^\omega}. \tag{15a}$$

Thus, the modes of the expansion of the position of the polymer chain segment are orthogonal and connected with the acting force by the relation

$$y_{q\alpha}^\omega = \frac{f_{q\alpha}^\omega}{-i\omega \lambda_q^\omega + \gamma_q},\quad \gamma_q = \frac{6\pi^2 k_B T}{Na^2} q^2. \tag{16}$$

Let us introduce the time correlation functions of the normal modes,

$$\psi_q(t) = \langle y_{q\alpha}(t) y_{q\alpha}(0) \rangle, \tag{17}$$

where the angular brackets denote the statistical average over the realization of the random forces. For the spectral density of this correlation function, $\psi_q^\omega$, with the use of the fluctuation-dissipation theorem [5], we find

$$\psi_q^\omega = \frac{k_B T}{\pi \omega} \operatorname{Im} \frac{1}{-i\omega \lambda_q^\omega + \gamma_q} = \operatorname{Re} \frac{k_B T}{\pi \gamma_q} \frac{\lambda_q^\omega}{-i\omega \lambda_q^\omega + \gamma_q}. \tag{18}$$

Taking into account the dependence of $\lambda_q^\omega$ from Eq. (15) on the frequency, $\psi_q^\omega$ can be expressed in the form of a proper fraction that in the numerator has a polynomial of the fourth degree and in the denominator a polynomial of the sixth degree in $\sqrt{-i\omega}$. Let $-\alpha_l$ be the simple roots of the polynomial in the denominator. In this case we shall have after the expansion into the simplest fractions

$$\psi_q^\omega = \operatorname{Re} \sum_{l=1}^{6} \frac{A_l}{\sqrt{-i\omega} + \alpha_l} = \operatorname{Re} \sum_{l=1}^{6} A_l \left( \frac{1}{\sqrt{-i\omega}} - \frac{\alpha_l}{\sqrt{-i\omega}\left(\sqrt{-i\omega} + \alpha_l\right)} \right). \tag{19}$$

Here $A_l$ are the expansion coefficients that can be easily determined explicitly. Using the result (18) one can prove that $\sum_l A_l = 0$. Returning into the $t$-representation we find

$$\psi_q(t) = -\pi \operatorname{Re} \sum_{l=1}^{6} A_l \alpha_l w\left(i\alpha_l \sqrt{t}\right), \tag{20}$$

where $w(z)$ is the Kramp function on the complex plane studied in detail in the work [9]. From this expression one immediately finds $\psi_q(t=0) = k_B T / \gamma_q$. The asymptotic behavior of the function $w(z)$ at large absolute values of its argument allows us to find the main contribution to the long-time correlation function of the normal modes,

$$\psi_q(t) \approx -\operatorname{Re} i \sum_{k \geq 0} \Gamma\left(k + \frac{1}{2}\right) \sum_{l=1}^{6} A_l \alpha_l \left(\frac{1}{i\alpha_l \sqrt{t}}\right)^{2k+1}. \tag{21}$$

As above, using the explicit expression for the spectral density (18), we find $\sum_l A_l / \alpha_l^2 = 0$, so that the first nonvanishing contribution is of a fractional power character as distinct from the traditional exponential one:

$$\psi_q(t) = -\operatorname{Re} \Gamma\left(\frac{5}{2}\right) \sum_{l=1}^{6} \frac{A_l}{\alpha_l^4} \left(\frac{1}{t}\right)^{5/2}. \tag{22}$$

The sum over the roots $\alpha_l$ is again easily calculated from the explicit form of the spectrum. Thus the viscous memory essentially influences the time dependence of the correlation functions of the Fourier amplitudes of the polymer chain fragments.

Since the radius vector of the center of inertia of the chain is

$$\vec{y}_0(t) = \frac{1}{N} \int_0^N \vec{x}(t,n) dn,$$

the diffusion coefficient of the polymer as a whole can be determined by the relation

$$D_c = \lim_{t\to\infty} \frac{1}{6t}\left\langle [\vec{y}_0(t) - \vec{y}_0(0)]^2 \right\rangle = \lim_{t\to\infty} \frac{1}{t}\int_{-\infty}^{\infty} d\omega (1-\cos\omega t)\psi_{q=0}^{\omega},$$

where in the second equality the Fourier transformation of Eq. (17) has been used. The integration with the function (15a) after the limit transition leads to the result

$$D_c = \frac{4k_B T}{3\eta a\sqrt{6\pi^3 N}},$$

that fully coincides with the expression for the diffusion coefficient of the polymer coil within the Zimm model. Finally, the relaxation time of the $q$th mode can be in our approach determined by the integral

$$\tau_q = \int_0^{\infty} \frac{\psi_q(t)}{\psi_q(0)} dt = \frac{(a\sqrt{N})^3 \eta}{k_B T q\sqrt{3\pi q}},$$

and it also agrees with the result based on the Zimm theory [1,2].